\documentclass[draftclsnofoot,onecolumn,11pt]{IEEEtran} 
 
\newcommand{\size}{N} 
\newcommand{\msize}{N}

\newcommand{\csize}{N^3}

\newcommand{\radius}{\rho}

\newcommand\given[1][]{\:#1\vert\:}

\usepackage{mathrsfs}

\newenvironment{definition}[1][Definition]{\begin{trivlist}
\item[\hskip \labelsep {\bfseries #1}]}{\end{trivlist}}

\newcommand{\qed}{\nobreak \ifvmode \relax \else 
      \ifdim\lastskip<1.5em \hskip-\lastskip
      \hskip1.5em plus0em minus0.5em \fi \nobreak
      \vrule height0.75em width0.5em depth0.25em\fi} 
      
\usepackage[backend=bibtex,style=ieee,sorting=none]{biblatex}
\bibliography{Refs}

\usepackage{graphicx}
\graphicspath{{figs/}}

\usepackage{amssymb}
\usepackage{bm}
\usepackage{commath}

\usepackage{amsthm}

\usepackage{enumitem}

\newtheorem{theorem}{\textbf{Theorem}}[section]
\newtheorem{theorem*}{\textbf{Theorem}}
\newtheorem{proposition*}{\textbf{Proposition}}

\newtheorem{lemma*}{\textbf{Lemma}}
\newtheorem{conjecture*}{\textbf{Conjecture}}

\newtheorem{corollary}[theorem]{Corollary}
\newtheorem{corollary*}{Corollary}

\begin{document}

\title{
Improved Intolerance Intervals and Size Bounds for a Schelling-Type Spin System 
 }



 \author{Hamed Omidvar, 
Massimo Franceschetti \\
Department of Electrical and Computer Engineering, \\University of California, San Diego \\
\texttt{ \{homidvar, mfrances\}@ucsd.edu}}

\maketitle

\begin{abstract}
We consider a Schelling model of self-organized segregation in an open system that is equivalent to a zero-temperature Ising model with Glauber dynamics, or an Asynchronous Cellular Automaton (ACA) with extended Moore neighborhoods. Previous work has shown that if the intolerance parameter of the model  $\tau\in (\sim 0.488,\sim 0.512) \setminus \{1/2\}$, then for a sufficiently large neighborhood of interaction $N$, any particle will end up in an exponentially large monochromatic region almost surely.
  
This paper extends the above result to the interval $\tau \in (\sim 0.433,\sim 0.567) \setminus \{1/2\}$. We also  improve the bounds on the size of the monochromatic region by exponential factors in $N$.
Finally, we show that when particles are placed on the infinite lattice $\mathbb{Z}^2$ rather than on a flat torus, for the values of $\tau$ mentioned above,  sufficiently large $N$, and after a sufficiently long evolution time,  any particle is contained in a large monochromatic region of size exponential in $N$, almost surely. The new proof, critically relies on a novel geometric construction related to the formation of the monochromatic region.
\end{abstract}

\section{Introduction}
 Consider the graph formed by nodes placed  at  the integer points of a large flat torus and edges connecting each node to all the ones located in a  square neighborhood of itself that is small compared to the size of the torus.  Put a particle at each node such that its initial binary state is chosen independently and uniformly at random. Each particle is then labeled as follows. All particles have a common intolerance threshold $0< \tau < 1$, indicating the minimum fraction of particles in their same state that must be in their neighborhood to make them ``stable." Each particle is assigned an independent and identical Poisson clock,  and when the clock rings  the particle's state is flipped if the particle is ``unstable" and the flip will make it stable. This change is then immediately detected by the neighbors   who update their labels accordingly.

This dynamical system model is known in the social sciences as the Schelling model in an ``open" system \cite{schelling1969models,schelling1971dynamic}.  In computation theory it corresponds to a two-dimensional, two-state Asynchronous Cellular Automaton (ACA) with extended Moore neighborhoods and exponential waiting times \cite{chopard1998cellular}.   For an intolerance value of $1/2$, the model corresponds to the Ising model with zero temperature, which exhibits spontaneous magnetization as spins align along the direction of the local field~\cite{stauffer2007ising, castellano2009}. 
Mathematically, the model falls into the  broad class of  interacting particle systems~\cite{liggett2012interacting,liggett2013stochastic}. 

The \textit{Glauber dynamics} we consider, assumes unstable particles to simply flip their state if this makes them stable. This flipping action indicates that the particle has moved out of the system and a new particle has occupied its location.  Alternatively,  \textit{Kawasaki dynamics} assume that pairs of unstable particles swap their locations if this will make both of them stable. The Kawasaki dynamics correspond to a ``closed" system where the number of particles in each state is fixed, while the Glauber dynamics correspond to an ``open" system where the number of particles in each state can change over time. 
Other variants are possible, 
including  having unstable particles swap (or flip) regardless of whether this makes them stable or not, or to assume that particles have a small probability of acting differently than what the general rule prescribes, 
have multiple intolerance levels,  multiple states, different distributions, and time-varying intolerance  \cite{young2001individual, zhang2004dynamic, zhang2004residential, zhang2011tipping, mobius2000formation, meyer2003immigration,bhakta2014clustering, schulze2005potts,barmpalias2015minority,barmpalias2015tipping}.


A common effect observed by simulating several variants of the model is that when the system reaches a steady state, large monochromatic areas of particles with the same state are formed, for a wide range of the intolerance threshold. This corresponds to observing spontaneous self-organization resulting from local interactions. 
%
%
Although simulations of this behavior have been available for a long time, rigorous results for the Schelling model appeared only  recently.  

The one-dimensional version of the model was studied rigorously by 
Brandt et al.~\cite{brandt2012analysis},  
Barmpalias et al.~\cite{barmpalias2014digital},  
and Holden and Sheffield  \cite{holden2018scaling}. 
In the two-dimensional model case,  Immorlica et al.~\cite{immorlica2015exponential}  have shown for the Glauber dynamics  that for  $\tau \in (1/2-\epsilon, 1/2)$ the  expected size of the monochromatic region   is exponential in the size of the neighborhood.  This $\epsilon$-size interval  has been enlarged in a previous work by the authors \cite{omidvar2017self2} to  $0.433 < \tau < 1/2$ (and by symmetry $1/2<\tau<0.567$).  In  the same work, the interval   has been further extended   to $0.344 < \tau \leq 0.433$   (and by symmetry for $0.567 \leq \tau<0.656$)  considering ``almost monochromatic''  regions, namely regions where the ratio of the  number of particles in one state and the  number of particles in the other state quickly vanishes as the size of the neighborhood grows. 
Barmpalias et al.~\cite{barmpalias2016unperturbed} considered a model in which particles in different states have different intolerance parameters, i.e.,  $\tau_1$ and $\tau_2$.  For the special case of $\tau_1 = \tau_2 = \tau$, they have shown that when  $\tau>3/4$, or  $\tau<1/4$,  the initial configuration remains static  w.h.p.
 
 In a   recent work by the authors \cite{omidvar2018shape}, a shape theorem for the spread of ``affected" nodes -- namely nodes on which a particle would be unstable in exactly one of its states --   has been provided.
This theorem gives a precise geometrical description of the set of affected nodes at any given time of the system evolution, and is a consequence of a concentration bound that was developed  for the spreading time. A size theorem has also been provided,  showing  the formation of exponential regions almost surely (a.s.), rather than just in expectation.  Namely,  for all  ${\tau \in (\tau^*,1-\tau^*) \setminus \{1/2\}}$ where ${\tau^* \approx 0.488}$, and when the size of the neighborhood of interaction $\size$ is sufficiently large,  any particle is contained in a large ``monochromatic region'' of size exponential in~$\size$  a.s. 

The main contributions of this paper are as follows.
First, we extend the  intolerance interval for the a.s. formation of exponential monochromatic regions to $\tau \in (\sim 0.433,\sim 0.567) \setminus \{1/2\}$. Second, we improve the bounds on the size of the monochromatic region by exponential factors.
Finally, we show that when particles are placed on the infinite lattice $\mathbb{Z}^2$ rather than on a flat torus, for the values of $\tau$ mentioned above,  sufficiently large $N$,   and after  a sufficiently long evolution time,  any particle is contained in a large monochromatic region of size exponential in $N$, almost surely.


 
In our model, the size of the neighborhood grows with $N$, leading to a long-range interaction model. Other works have considered the case of constant neighborhood.
Fontes et al.~\cite{siv} have shown the existence of a critical probability  $1/2<p^*<1$ for the initial Bernoulli distribution of the particle states such that for $\tau =1/2$  and $p > p^*$ the Glauber model on the $d$-dimensional grid converges to a state where only particles in one state are present. This shows that complete monochromaticity occurs w.h.p.\ for $\tau=1/2$ and $p \in (1-\epsilon,1)$. Morris~\cite{morris2011zero} has shown that $p^*$   converges to $1/2$ as $d \rightarrow \infty$. Caputo and Martinelli \cite{caputo2006phase} have shown the same result for $d$-regular trees, while  Kanoria and Montanari~\cite{montanaritree} derived it for  $d$-regular trees in a synchronous setting where  flips occur simultaneously,  and obtained lower bounds on $p^*(d)$ for small values of $d$. The case $d = 1$   was first investigated by Erd\"{o}s and Ney~\cite{erdos1974some}, and Arratia~\cite{arratia1983site} has proven that $p^*(1)=1$.

The  rest of the paper is organized as follows. In section \ref{Sec:Model} we introduce the model, state our results, and give a summary of the proof construction.  
In section~\ref{Sec:Prelim} we provide a few preliminary results along with some results from previous works.
In section~\ref{Sec:Proof} we prove Theorem~1.

\section{Model and Main Results}\label{Sec:Model}
\subsection{The Model} 
\textit{Initial Configuration.}
Consider a node at each integer point of a flat torus $\mathbb{T}=[-h,h) \times [-h,h)$ where $h\in \mathbb{N}$ and let the \textit{horizon} $w \in o( \sqrt{\log h})$ be a natural number such that each node is connected to all the nodes located within an $l_\infty$ neighborhood of radius $w$ of itself. We place a particle at each node of the resulting graph $G_w$. Each particle has a binary state which is chosen independently at random to be (+1) or (-1) according to a Bernoulli distribution of parameter $p  = 1/2$. We use $\theta$ to denote an unknown or unspecified state and $\bar{\theta}$ to denote its complement state and by a $\theta$-particle we mean a particle that is in state $\theta$.

\

A \textit{neighborhood} is a connected sub-graph of $G_w$. 
A  \textit{neighborhood of radius ${\radius}$}, denoted by $\rho$-neighborhood or $\mathcal{N}_{\radius}$, is the set of all particles with $l_\infty$ distance at most ${\radius}$ from a central node. The \textit{size} of a neighborhood is the  number of particles in  it.   
The \textit{neighborhood of a particle $u$} is a neighborhood of radius equal to the horizon and centered at $u$, and is denoted by $\mathcal{N}(u)$. Without loss of generality, we derive our result for an arbitrary particle located at the origin  $0 \equiv (0,0)\in \mathbb{T}$.

\

\textit{Dynamics.}
 We let the rational $\tau$, called \emph{intolerance}, be $\lceil \tilde{\tau} {\size} \rceil/{\size}$, where $\tilde{\tau} \in [0,1]$   and 
${\size} = (2w+1)^2$  is the size of the neighborhood of a particle. The integer $\tau {\size}$ represents the minimum number of particles in the same state as $u$ that must be present in $\mathcal{N}(u)$  to make   $u$  \emph{stable}. More precisely,
for every particle $u$, we let $s(u)$ be the ratio between the number of particles that are in the same state as $u$ in its neighborhood and the size of the neighborhood. 
At any point in continuous-time, if ${s(u) \ge \tau}$ then $u$ is labeled \textit{stable}, otherwise it is labeled \textit{unstable}. We assign independent and identically distributed waiting times to unstable particles, and after every waiting time, the state of the particle is flipped if and only if the particle is still unstable and this flip will make the particle stable. Since we consider Glauber dynamics this means $F$ is exponential, however, all of our results hold
for distributions that satisfy the following conditions:
\begin{align} \label{eq:1.1}
F(x) = 0 \mbox{ for } x\le 0,\\
F \text{ is not concentrated on one point,} \label{eq:1.2}
\end{align}
 and
\begin{align}\label{eq:1.14}
\exists \gamma >0, \mbox{ such that } \int e^{\gamma x} F(dx) < \infty. 
\end{align}

 Two observations are now  in order. First, for $\tau< 1/2$  flipping its state will always make an unstable particle stable,  but this is not the case  for $\tau>1/2$. Second, the process dynamics for the case of having exponential distributions are equivalent to a discrete-time model where at each discrete time step one unstable particle is chosen uniformly at random and its state is flipped if this will make the particle stable.

\

\textit{Affected Nodes.} A node in $G_w$ is called $\theta$-\textit{affected} whenever a $\theta$-particle located there would be unstable and a $\bar{\theta}$-particle would be stable.

\

\textit{Steady State.}
The process continues until there are no unstable particles left, or there are no unstable particles that can become stable by flipping their state. 
By defining a Lyapunov function to be  the sum over all particles $u$ of the number of  particles of the same state as $u$ present in its neighborhood, it is easy to argue that the process indeed terminates. We call this final state the \textit{steady state}.

\

\textit{Monochromatic Region.}
At each point in time, the \textit{monochromatic regions} of a particle $u$  are the neighborhoods with largest radius that contain particles in a single state and  that  also  contain $u$. The \textit{monochromatic region} of $u$ is  one of these regions chosen arbitrarily.

\

Throughout the paper we use the terminology \textit{with high probability} (w.h.p.) meaning that the probability of an event is $1-o(w^{-2})$ as $w \rightarrow \infty$. 
Almost surely (a.s.)  instead indicates \textit{with probability equal to one} (w.p.1). 

\subsection{Main Results}

To state the results, 
we let $\epsilon \in (0,0.1)$ , and  $\tau^* \approx 0.433$ be the minimum value of $\tau$ that satisfies
\begin{align}
\frac{(3/2+g(\tau))^2\tau}{4}{\size} + \frac{1}{2}\left(1-\frac{(3/2+g(\tau))^2}{4}\right){\size} - \tau(0.265)^2{\size} < \tau {\size},\\
\frac{3}{4} \left(1-H\left(\frac{4}{3} \tau_* \right)\right)- \max\{\frac{3}{2}\log_2(\rho),2\log_2(\rho')\}< 0,\\
0.5\log_2 (\rho)-\log_2 \rho' < 0, \\
\log_2(\rho'')-\log_2(\rho') < 0,
\end{align}
where 
\begin{align}
g(\tau) = \frac{(694\tau + 800(((2\tau - 1)(409\tau - 20000))/80000)^{1/2} - 347)}{200(6\tau + 1)} , \label{eq:g} \\
\rho =2^{0.5(1-H(\tau)+\epsilon/2)(1+g(\tau))^2\size}, \\
\rho' = 2^{(1-H(\tau)-\epsilon)(1-0.5(1+g(\tau))^2)\size + 2\log_2 \size},\\
\rho'' = 2^{(1-H(\tau)+\epsilon)\left((1+g(\tau))^2-1\right)\size}, \\
H(\tau) = -\tau \log_2(\tau) - (1-\tau)\log_2(1-\tau).
\label{eq:ftau}
\end{align}



\begin{theorem*}[Size Theorem ---Steady State]\label{Thrm:main_thrm}
Let $M$ denote the size of the monochromatic region of the particle at the origin in the steady state. 
For all $\tau \in (\tau^*,1-\tau^*) \setminus \{1/2\}$ and for sufficiently large $\size$, almost surely,
\begin{align}
 2^{a(\tau){\size}} \le M \le 2^{b(\tau){\size}}.
\end{align}
\end{theorem*}

\begin{figure}
\centering
\includegraphics[width=3.5in]{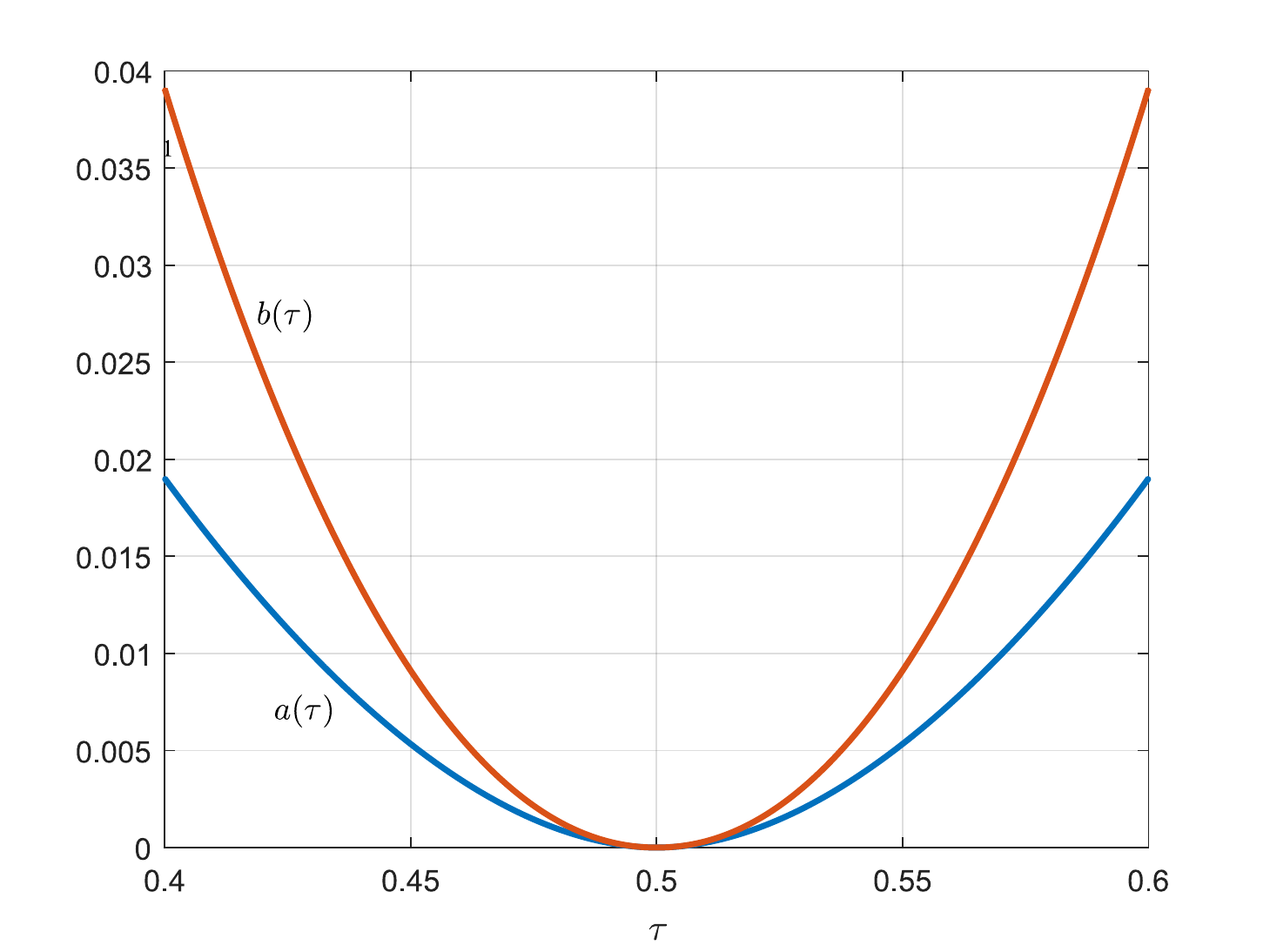}
\caption{Exponent multipliers $a(\tau)$ and $b(\tau)$ for the lower bound and upper bounds on the size of the monochromatic region.} 
\label{fig:aaprime}
\end{figure}

The numerical values for $a(\tau)$ and $b(\tau)$  derived in  the proof of the above theorem are plotted in Figure~\ref{fig:aaprime}. For $\tau \in (\tau^*,1-\tau^*)\setminus \{1/2\}$,  as the intolerance gets farther from one half in both directions, larger monochromatic regions are formed almost surely.

\subsection{The Infinite Lattice Case}
We can consider similar dynamics occurring on the infinite lattice $\mathbb{Z}^2$ instead of the finite setting described above. 
 Using Theorem B3 of \cite[p.~3]{liggett2013stochastic} it is easy to verify that  the process on  $\mathbb{Z}^2$   exists, and is unique, and is a Feller Markov process on $\{-1,+1\}^{\mathbb{Z}^2}$.
The following corollary follows from the proof of Theorem~\ref{Thrm:main_thrm}.

\begin{corollary}[Size Theorem ---Infinite Lattice]\label{Corr:size_thrm}
Let $M_n$ denote the size of the monochromatic region of the particle at the origin at time $n$. For all $\epsilon>0$, $\tau \in (\tau^*,1-\tau^*) \setminus \{1/2\}$, let $n^* = 2^{{(a(\tau)+\epsilon)}\size}$. For sufficiently large $\size$, and all $n \ge n^*$, almost surely,
\begin{align}
 2^{a(\tau){\size}} \le M_n \le  2^{b(\tau){\size}}.
\end{align}
\end{corollary}


\subsection{Proof Outline}
 \begin{figure}[!t] 
\centering
\includegraphics[width=3.7in]{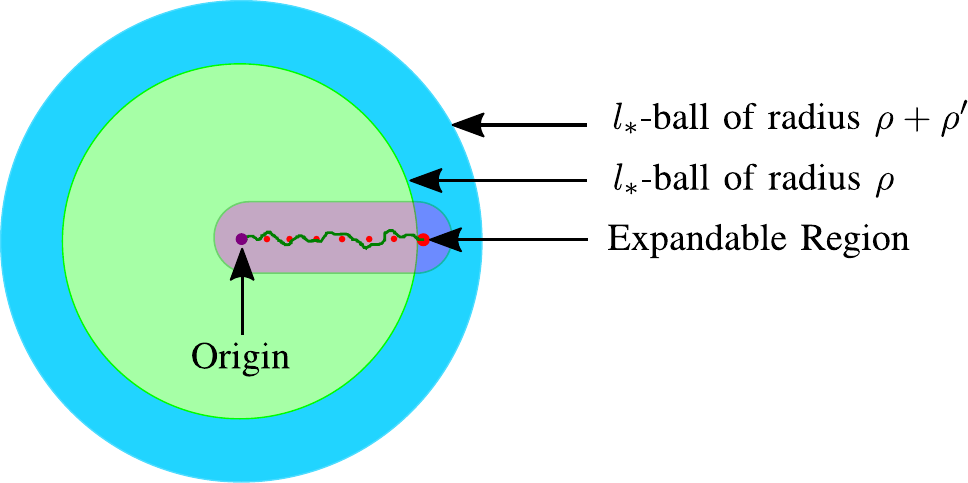}
\caption{The gradual spread of affected nodes from the expandable region towards the origin.} 
\label{fig:gradual_growth}
\end{figure} 

The general strategy of the proof follows the one provided in \cite{omidvar2018shape} -- we show that while the spread of the $\theta$-affected nodes (i.e., nodes on which a $\theta$-particle would be unstable) reaches the origin, the $\bar{\theta}$-affected nodes are still at distances at least exponential in $\size$ from the origin. 
 Once the origin is reached, the unstable particles around it  will w.h.p. lead to the formation of an exponentially large ``firewall"   that is indestructible by other spreading processes. The interior of this firewall will then become monochromatic, so that in the steady state there will be  w.h.p. an exponentially large monochromatic region around the origin.
 
 The main difference in this proof compared to the one in \cite{omidvar2018shape}, however, is that instead of considering the spreading of affected nodes towards the origin in multiple stages -- which leads to a large aggregated error term -- we consider the optimal sequence of flips reaching the origin in one shot. This is possible thanks to improved bounds and geometrical constructions that we provide throughout this paper. 

To elaborate on the main idea, we define an \textit{expandable region}, that  is composed of a local configuration of particles and a possible set of flips inside it, that can lead to at least one new affected node outside of it.  We consider the expandable region closest to the origin in a norm $l_*$ -- whose existence is proved using results from \cite{omidvar2018shape} and is used for describing the spread of affected nodes -- and denote its type by $\theta$. We denote its $l_*$-distance to the origin by $X$, and  consider an $l_*$-ball of radius $X$ at the origin. We then argue that, since there are no expandable regions in this ball, any spreading of affected nodes inside this ball dies out quickly, while  the expandable region starts a spreading of $\theta$-affected nodes towards the origin w.h.p. 
We then find an upper bound $X \leq \rho$, where $\rho=\rho(\tau,\size)$, that holds w.h.p.,  and choose $\rho'=\rho'(\tau,\size)$ such that  w.h.p there is no $\bar{\theta}$-expandable region inside the annulus $B_{l_*}(0,X+\rho') \setminus B_{l_*}(0,X)$.
We  consider the worst case $X=\rho$ and argue that the possible spreads of the $\bar{\theta}$-affected nodes from outside $B_{l_*}(0,\rho+\rho')$ are not going to be capable of disrupting the spread of the expandable region at distance $\rho$ towards the origin,  see Figure~\ref{fig:gradual_growth}. 

\begin{figure}[!t] 
\centering
\includegraphics[width=1.6in]{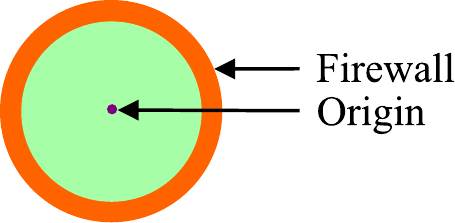}
\caption{A firewall formed around the origin.} 
\label{fig:fw0}
\end{figure}

Instead of considering gradual spreads of the expandable region in separate time intervals of $\rho'/4$ towards the origin as in~\cite{omidvar2018shape}, we argue that the optimal sequence of flips reaching the origin is going to be close to a straight line connecting the two points and these flips are not going to be affected by the spread of $\bar{\theta}$-affected nodes (by looking at this one shot growth in $\rho'/4$ time intervals). We then show that  the origin is quickly surrounded by  an exponentially large \textit{firewall} while any spreading of affected nodes started from outside $B_{l_*}(0,\rho+\rho')$  is still at large distances from it. This firewall is an indestructible monochromatic annulus which isolates the origin from the outside flips, see Figure~\ref{fig:fw0}. It  will thus protect the cascading process which w.h.p. leads to the formation of a monochromatic region of size exponential in $\size$ containing the origin. This shows that the lower bound occurs w.h.p. 
To see that the upper bound also occurs w.h.p. we note that in a large enough exponential size neighborhood around the origin, w.h.p. the origin will be surrounded by exponentially large monochromatic regions of particles in both states protected by firewalls. 
Finally, we let $A_w$ for $w=1,2,...$ be the sequence of events of having the origin contained in monochromatic regions of size exponential in $\size$ in an appropriate probability space. We note that these events occur w.h.p. (so their complements occur with probability $o(w^{-2})$). The proof now follows from the fact that $\sum_w w^{-2} < \infty$ using the Borel-Cantelli lemma.
\

In our proofs throughout the paper, we focus on the case where $\tau < 1/2$. The results for $\tau > 1/2$ follow by a simple symmetry argument provided in Section~\ref{Subsec:Extension}.

\section{Preliminary and Previous Results}\label{Sec:Prelim}
We begin with the following elementary lemma giving lower and upper bounds for the probability of a node being affected. 

\begin{lemma*}[\cite{omidvar2018shape}]\label{lemma:unstableprob}
Let $p_u$ be the probability of being $\theta$-affected for an arbitrary node in the initial configuration. There exist positive constants $c_l$ and $c_u$ which depend only on $\tau$ such that
\begin{align*}
c_{l}\frac{2^{-[1-H(\tau')]{\size}}}{\sqrt{{\size}}}  \le p_u \le c_{u}\frac{2^{-[1-H(\tau')]{\size}}}{\sqrt{{\size}}},
\end{align*}
 where $\tau' = \frac{\tau {\size} - 2}{{\size}-1}$, and $H(\tau)= -\tau \log_2(\tau) - (1-\tau)\log_2(1-\tau)$ is the binary entropy function.
\end{lemma*}

The following simple lemma is a consequence of Lemma~\ref{lemma:unstableprob}.
\begin{lemma*}[\cite{omidvar2018shape}] \label{lemma:R_unstable}
Let ${\radius}  =  2^{[1-H(\tau')]{\size}/2}$. The following event occurs w.h.p. 
\begin{align*}
A = \left\{\nexists \  \theta\mbox{-affected node in } \mathcal{N}_{\rho}\right\}. 
\end{align*}
\end{lemma*}

\textit{$m$-block.} We define an $m$-\textit{block} to be a neighborhood of radius $m/2$. A \textit{monochromatic block} is a block whose particles are all at the same state. When $m$ is not specified, by a \textit{block} we mean a $w$-block.

\

\textit{Region of expansion.} We call  a \emph{region of expansion} (of type $\theta$) any neighborhood whose configuration is such that by placing a monochromatic $w$-block with all particles in state $\bar{\theta}$ anywhere inside it, all the $\theta$-particles on its outside boundary (i.e., the set of particles in the set $(w+2)$-block co-centered with the $w$-block excluding the $w$-block itself) become unstable w.p.1. By a region of expansion we mean a region of expansion of only one type ($\theta$). The next lemma, which is a restatement of Lemma~8 in \cite{omidvar2017self2}, shows that as long as $\tau \in (\tau_*,1/2)$, w.h.p. a monochromatic block on $G_w$ can make an exponentially large area monochromatic.

\begin{lemma*}[\cite{omidvar2017self2}]  \label{lemma:monoch_spread_1}
Let $\tau \in (\tau_*,1/2)$ and let $\mathcal{N}_r$ be a neighborhood with radius $r<2^{0.5{[1-H(\tau')]{\size}-o({\size})}}$ in the initial configuration. $\mathcal{N}_r$ is a region of expansion w.h.p.
\end{lemma*}

Let $\mathcal{I}_x$ be the collection of sets of particles  in $w/2$-neighborhoods in an $m$-block ($m\ge w$) on $G_w$  in the initial configuration. Also, let $W_{I}$ be the random variable representing the number of particles in state $\bar{\theta}$ in $I \in \mathcal{I}_x$, and ${\size}_I$ be the total number of particles in $I \in \mathcal{I}_x$.

\begin{figure}[!t]
\centering
\includegraphics[width=2in]{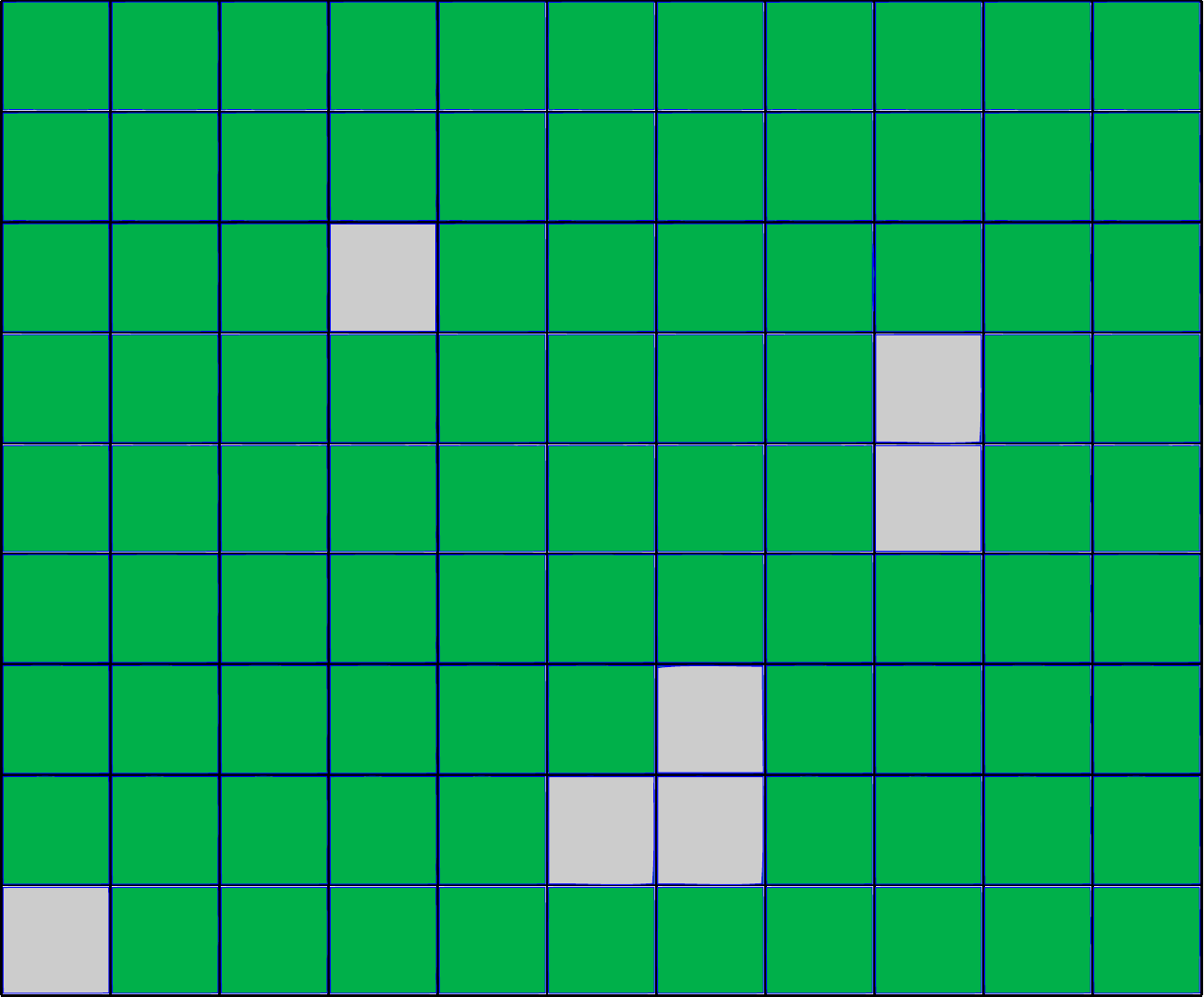}
\caption{Part of the lattice renormalized into $m$-blocks.  Green and gray indicate good and bad blocks respectively. }
\label{fig:goodbad}
\end{figure}

\textit{Good block.} For any $\epsilon \in (0,1/2)$, a \textit{good $m$-block} of type $\theta$ is an $m$-block, such that  for all $I\in \mathcal{I}_x$ we have $W_I-{\size}_I/2  < {\size}^{1/2+\epsilon}$. An $m$-block that does not satisfy this property is called a \textit{bad $m$-block} (see Figure~\ref{fig:goodbad}). By the following lemma (which is a restatement of Lemma~11 in~\cite{omidvar2017self2}), an $\msize$-block is a good block w.h.p. Since the number of different particles in a good block is balanced enough for sufficiently large $\size$, a node whose entire neighborhood is contained in a good block cannot be a $\theta$-affected node (we assume throughout that $\size$ is large enough such that this is the case).

\begin{lemma*}[\cite{omidvar2017self2}] \label{lemma:goodblock}
Let $\epsilon \in (0,1/2)$. There exists a constant $c>0$, such that, for all $I\in \mathcal{I}$ we have
\begin{align*} 
W_I-{\size}_I/2 < {\size}^{1/2+\epsilon}
\end{align*} 
with probability at least
\begin{align*}
1 - e^{-c{\size}^{2\epsilon}+o({\size}^{2\epsilon})}.
\end{align*} 
\end{lemma*}

We now want to review a result from percolation theory. Without loss of generality, we assume $G_w$ is defined on $\mathbb{Z}^2$. Let us re-normalize $G_w$ into $m$-blocks. Let $S(k)$ be the ball of radius $k$ with center at the origin, i.e., $S(k)$ is the set of all $m$-blocks $x$ on the re-normalized $G_w$ for which $\Delta(0,x) \le k$, where $\Delta$ denotes the $l_\infty$ distance on the re-normalized $G_w$. Let $\partial S(k)$ denote the surface of $S(k)$, i.e., the set of all $x$ such that $\Delta(0,x)=k$. We define a path of $m$-blocks as an ordered set of $m$-blocks such that each pair of consecutive $m$-blocks are in the Moore neighborhood (set of nearest $l_\infty$-neighbors) of each other and no $m$-block appears more than once in the set. Let $A_k$ be the event that there exists a path of bad $m$-blocks joining the origin to some vertex in $\partial S(k)$. Let the \textit{radius} of a bad cluster (i.e., cluster of bad $m$-blocks) be defined as 
\begin{align*}
\sup \{\Delta(0,x): x \in \mbox{bad cluster}\}.
\end{align*}
Let $p$ denote the probaility of having a bad $m$-block.  It is noted that an $m$-block is a bad $m$-block independently of the others. Let $p_c$ denote the ciritical probability in the above percolation setting. The following result is Theorem 5.4 in \cite{grimmett1999percolation}.

\begin{theorem*}\label{Thrm:grimmett_bad_cluster} 
 \textsl{(Exponential tail decay of the radius of a bad cluster.)}
If $p<p_c$, there exists $\psi(p) >0 $ such that
\begin{align*} 
P_p(A_k) < e^{-k\psi(p)}, \ \ \  \mbox{ for all } \ k.
\end{align*}
\end{theorem*}

\textit{Firewall.} A \textit{firewall} of radius $r$ and center $u$ is a monochromatic annulus
\begin{align*}
A_r(u) = \left\{y: r-\sqrt{2}w \leq \|u-y\|_2 \leq r\right\},
\end{align*}
    where $\|.\|_2$ denotes  Euclidean distance and $r\ge 3w$.

Consider a disc of radius $r$,   centered at a particle such that all the particles inside the disc are in the same state. It is easy to see that if $r$ is sufficiently large then all the particles inside the disc will remain stable regardless of the configuration of the particles outside the disc. Lemma~6 in \cite{immorlica2015exponential} shows that for $r>w^3$ this would be the case for sufficiently large $w$. Here we state a similar lemma but for a firewall, without proof.

\begin{lemma*}[\cite{immorlica2015exponential}]  \label{lemma:firewall}
Let $A_r(u)$ be the set of particles contained in an annulus of outer radius $r \ge w^3$ and of width $\sqrt{2} w$ centered at $u$. For all $\tau \in (\tau^*,1/2)$ and for a sufficiently large constant $\size$, if $A_r(u)$ is monochromatic at time $n$,  then it will remain monochromatic at all times  $n'>n$. 
\end{lemma*}

 By Lemma~\ref{lemma:firewall}, once formed a firewall of sufficiently large radius  remains  static,  and since its width is  $\sqrt{2} w$ the particles inside the inner circle are not going to be affected by the configurations outside the firewall.

\

We now review some results and definitions from \cite{omidvar2017self2}. In the current paper, the goal is to identify a better configuration, compared to the one constructed in \cite{omidvar2017self2}, that can trigger a cascading process leading to monochromatic regions w.h.p.  The following proposition is Proposition~1 in~\cite{omidvar2017self2}.

\

\begin{proposition*}[\cite{omidvar2017self2}]\label{Prop:firstprop}
For any $\epsilon \in (0,1/2)$ and $c \in \mathbb{R}^+$ there exists  $c' \in \mathbb{R}^{+}$  such that for all $\size \ge 1$
\begin{align*}
P\left(|W' -  \gamma \tau {\size}| <  c{\size}^{1/2+\epsilon} \given[\Big] W < \tau {\size}\right) \ge 1 - e^{-c'{\size}^{2\epsilon}}.
\end{align*}
\end{proposition*}

The following lemma, is Lemma~18 in \cite{omidvar2017self2}.

\begin{lemma*}[\cite{omidvar2017self2}]\label{lemma:balanced}
Let $\epsilon \in (0,1/2)$, and let $\mathcal{N}$ be an arbitrary neighborhood in the grid with ${\size}$ particles. 
There exist $c,c' \in \mathbb{R}^+$, such that 
\begin{align}\label{Eq:balanced}
P\left(|W - {\size}/2| < c{\size}^{1/2+\epsilon}\right) \ge 1-2e^{-c'{\size}^{2\epsilon}}.
\end{align}
\end{lemma*} 

\textit{Radical region.}
For any $\epsilon, \epsilon' \in {(0,1/2)}$ let ${\hat{\tau} = \tau [1- 1/ (\tau {\size}^{1/2-\epsilon})]}$ and define  a
\textit{radical region}  to be a neighborhood 
of radius ${(1+\epsilon')w}$ containing  less than $\hat{\tau}(1+\epsilon')^2{\size} $  particles in state $\theta$. 



 We define
an \textit{unstable region} to be a neighborhood of radius $\epsilon' w$, containing at least $\lfloor \tau \epsilon'^2 {\size}-{\size}^{1/2+\epsilon} \rfloor$ unstable particles in state $\theta$. 
The following is Lemma~4 in \cite{omidvar2017self2}.
\begin{lemma*} [\cite{omidvar2017self2}] \label{lemma:unstable_region}
A radical region $\mathcal{N}_{(1+\epsilon')w}$ in the initial configuration contains an unstable region $\mathcal{N}_{\epsilon'w}$ at its center w.h.p.
\end{lemma*}



Now consider a geometric configuration 
where a radical region,  and  neighborhoods $\mathcal{N}_{\epsilon'w}$ , $\mathcal{N}_{w/2}$ and $\mathcal{N}_{{\radius}}$  with ${\radius}>3w$, are all co-centered. 
Let
\begin{align}
T({\radius}) = \inf\{n: \exists v\in \mathcal{N}_{\radius}, \;  v \mbox{ is a $\bar{\theta}$-affected node} \}.
\label{Tinfdef}
\end{align}



\textit{Expandable radical region.} A radical region is called an \emph{expandable} radical region of type $\theta$ if there is a sequence of  at most $(w+1)^2$ possible flips inside it that can make  the neighborhood $\mathcal{N}_{w/2}$ at its center monochromatic with particles in state $\bar{\theta}$. The next lemma, shows that a radical region in this configuration is an expandable radical region of type $\theta$ w.h.p., provided that  $\epsilon'$ is large enough and there is no $\bar{\theta}$-affected node in $\mathcal{N}_{\radius}$.  The main idea is that the $\bar{\theta}$ particles in the unstable region at the center of the radical region can  trigger a process that leads to a monochromatic region of radius $w$.

\begin{lemma*} \label{lemma:Trigger}
For all $\epsilon' > g(\tau)$, where $g(\tau)$ is as defined in (\ref{eq:g}),
there exists w.h.p. a sequence of at most $(w+1)^2$ possible flips  in $\mathcal{N}_{(1+\epsilon')w}$ such that if they  happen  before $T({\radius})$,  then all the particles inside   $\mathcal{N}_{w/2}$ will become of  type $\bar{\theta}$.
\end{lemma*}

\begin{figure}[!t]
\centering
\includegraphics[width=\textwidth]{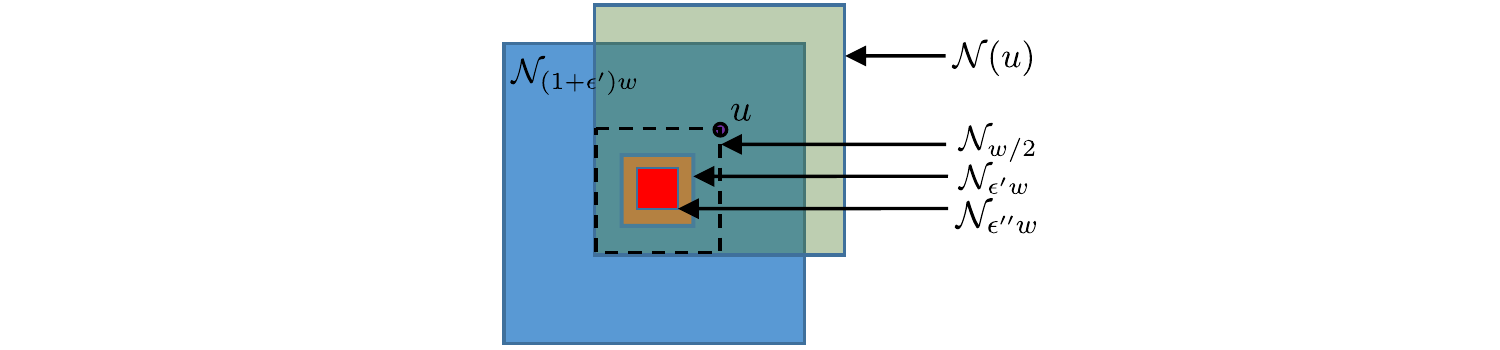}
\caption{Regions discussed in lemma~\ref{lemma:Trigger}. $\mathcal{N}_{\epsilon'w}$ is an unstable region w.h.p., the dashed box is $\mathcal{N}_{w/2}$, $u$ is a corner particle in $\mathcal{N}_{w/2}$, and finally $\mathcal{N}(u)$ is the neighborhood of particle $u$.}
\label{fig:radical}
\end{figure}

\begin{proof}
Let $\epsilon \in (0,1/2), \epsilon'' = 0.265$. Let us denote the neighborhoods with radii $\epsilon'w,\epsilon''w$ and co-centered with the radical region by $\mathcal{N}_{\epsilon'w},\mathcal{N}_{\epsilon''w}$ respectively (see Figure \ref{fig:radical}). By lemma \ref{lemma:unstable_region}, with probability at least $1-e^{-O({\size}^{2\epsilon})}$ there are at least $\lfloor \tau\epsilon'^2{\size} - {\size}^{2\epsilon} \rfloor$ particles in state $\theta$ inside this neighborhood such that all of them are unstable. Also using Proposition~\ref{Prop:firstprop} there are a total of at least $\lfloor \tau\epsilon''^2{\size} - {\size}^{2\epsilon} \rfloor$ unstable particles in $\mathcal{N}_{\epsilon''w}$.  Next, we  show that if the unstable particles in $\mathcal{N}_{\epsilon'w}$ flip before $T(\radius)$, all the particles inside the neighborhood $\mathcal{N}_{\epsilon''w}$ will be unstable w.h.p. and then if these unstable particles make a flip before $T(\radius)$ as well, then all the particles in $\mathcal{N}_{w/2}$ will be unstable w.h.p. which gives the desired result.

First, we notice that if there is a flip of an unstable $\theta$ particle in $\mathcal{N}_{\radius}\setminus \mathcal{N}_{w/2}$ it can only increase the probability of the existence of the sequence of flips we are looking for, hence conditioned on having these flips before $T(\radius)$, the worst case is when these flips occur with the initial configuration of $\mathcal{N}_{\radius}\setminus \mathcal{N}_{w/2}$. 
Since a corner particle in $\mathcal{N}_{\epsilon''w}$ shares the least number of particles with the radical region, it is more likely for it to have the largest number of $\bar{\theta}$ particles in its neighborhood compared to other particles in $\mathcal{N}_{\epsilon''w}$. Hence, as a worst case, we may consider a corner particle in $\mathcal{N}_{\epsilon''w}$ which is co-centered with the radical region.

Let us assume that $\epsilon' \in (0,0.265)$, in this case $\mathcal{N}_{\epsilon'w}$ is  completely contained in the neighborhood of each of the particles in   $\mathcal{N}_{\epsilon''w}$. Let us denote the neighborhood shared between the neighborhood of the particle $u$ at the corner of $\mathcal{N}_{\epsilon''w}$ and the radical region by $\mathcal{N}''(u)$. Also, let us denote the  scaling factor corresponding to this shared neighborhood by $\gamma''$. We have
 \begin{align*}
 \gamma''= \frac{(2+\epsilon'-\epsilon'')^2}{4(1+\epsilon')^2} \pm O\left(\frac{1}{\sqrt{{\size}}}\right).
 \end{align*}
By Proposition \ref{Prop:firstprop} it follows that with probability at least ${1-e^{-O({\size}^{2\epsilon})}}$ there are at most
  \begin{align*}
\frac{(2+\epsilon'-\epsilon'')^2\tau}{4}{\size} + o(N),
  \end{align*}
particles of type $\theta$ in $\mathcal{N}''(u)$. Hence, we can conclude that, for any particle in $\mathcal{N}_{w/2}$, w.h.p., there are at most this many $\theta$ particles in the intersection of the neighborhood of this particle and the radical region.

Also, using lemma \ref{lemma:balanced}, with probability at least $1-e^{-O(N^{2\epsilon})}$ we have at most
  \begin{align*}
  { \frac{1}{2}\left(1-(2+\epsilon'-\epsilon'')^2/4\right){\size} + o({\size}) },
  \end{align*}
 particles of type $\theta$ in the part of the neighborhood of the corner particle $u$  in $\mathcal{N}_{\epsilon''w}$ that is also not  in the radical region. Combining the above results, we can conclude that with probability at least $1-e^{-O(N^{2\epsilon})}$ there are at most 
\begin{align*}
\frac{(2+\epsilon'-\epsilon'')^2\tau}{4}{\size} + \frac{1}{2}\left(1-\frac{(2+\epsilon'-\epsilon'')^2\tau}{4}\right){\size}  + o({\size}),
 \end{align*} 
particles of type $\theta$ in the neighborhood of an particle in $\mathcal{N}_{w/2}$. Let us denote this event for the corner particle~$u$ by $A_1$. Let us denote the events of having at most this many $\theta$ particles in the neighborhoods of other particles in $\mathcal{N}_{w/2}$ by $A_2, ..., A_{|\mathcal{N}_{w/2}|}$, where $|\mathcal{N}_{\epsilon''w}|$ denotes the number of particles in $\mathcal{N}_{\epsilon''w}$. We have
\begin{align*}
P(A_1 \cap ... \cap A_{(w+1)^2}) &\ge 1 - P(A_1^C \cup ... \cup A^C_{|\mathcal{N}_{\epsilon''w}|}) \\
&\ge 1 - (w+1)^2 P(A^C_1) \\ 
& \ge 1 - e^{-O(N^{2\epsilon})}.
\end{align*}


The goal is now to find the range of $\epsilon'$ for which $\mathcal{N}_{\epsilon'w}$ is large enough that once all of its unstable particles flip, all the particles in $\mathcal{N}_{\epsilon''w}$ become unstable w.h.p. 
It follows that we need 
\begin{align*}
\frac{(2+\epsilon'-\epsilon'')^2\tau}{4}{\size} + \frac{1}{2}\left(1-\frac{(2+\epsilon'-\epsilon'')^2}{4}\right){\size} - \tau\epsilon'^2{\size} + o({\size}) < \tau {\size},
 \end{align*} 
to hold w.h.p. Dividing by ${\size}$, and letting ${\size}$ go to infinity,  after some   algebra it follows that
 \begin{align}\label{Eq:eps1}
\epsilon' > \frac{(694\tau + 800(((2\tau - 1)(409\tau - 20000))/80000)^{1/2} - 347)}{200(6\tau + 1)} = g(\tau),
\end{align}
where $g(\tau) < 0.265$ for $\tau \in (\tau^*,1/2)$, as desired. Now we need to make sure that once all the unstable particles in $\mathcal{N}_{\epsilon''w}$ make a flip, all the $\bar{\theta}$ particles in the $\mathcal{N}_{w/2}$ are going to be unstable. Similar to the  above argument, we need to have
\begin{align*}
\frac{(3/2+\epsilon')^2\tau}{4}{\size} + \frac{1}{2}\left(1-\frac{(3/2+\epsilon')^2}{4}\right){\size} - \tau\epsilon''^2{\size} + o({\size}) < \tau {\size},
 \end{align*} 
 which is true for $\tau \in (\tau^*,1/2)$, as desired.
\end{proof}

\begin{figure}[!t]
\centering
\includegraphics[width=4.5in]{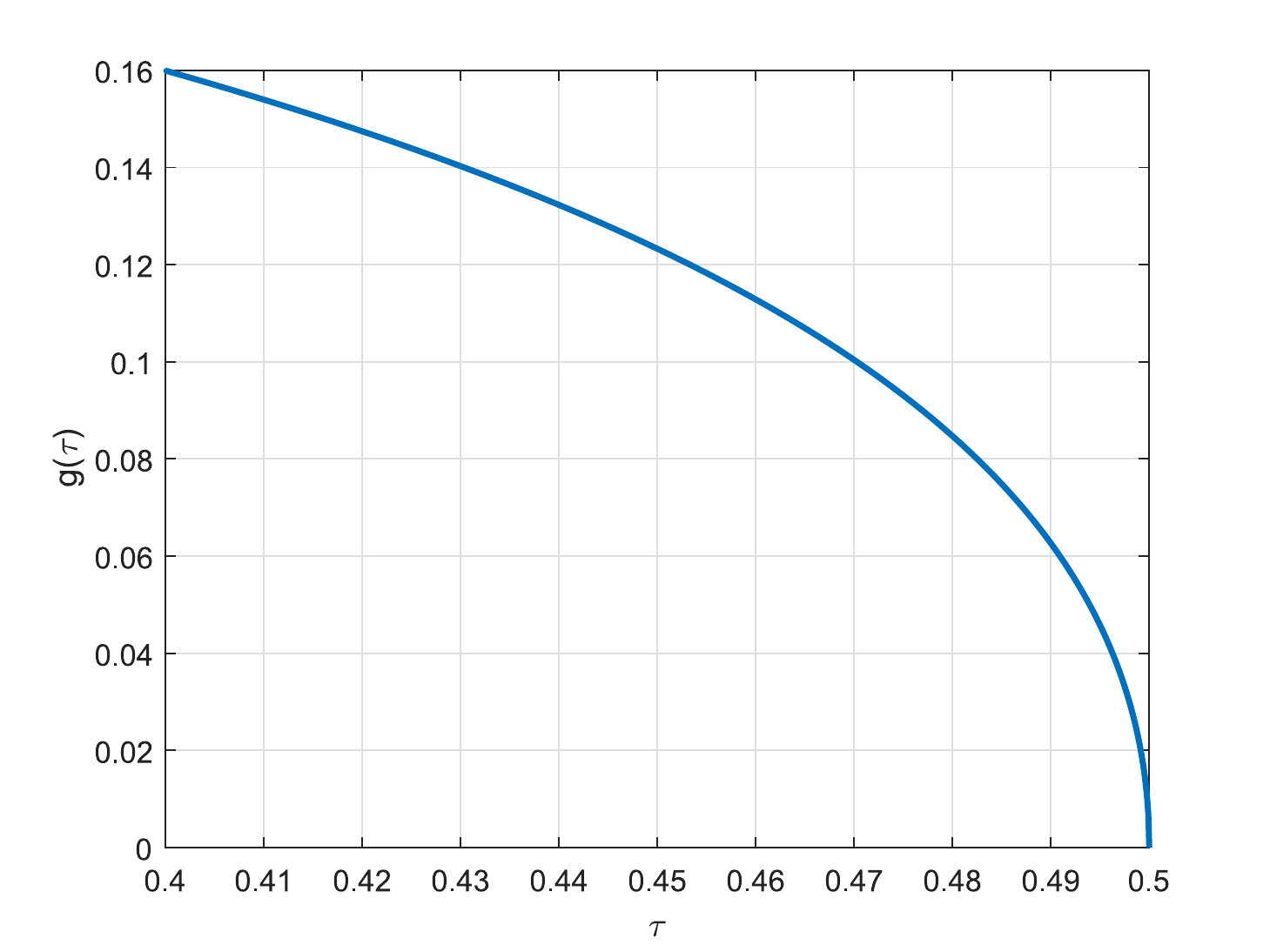}
\caption{The infimum of $\epsilon'$ to potentially trigger a cascading process. The value of $g(\tau)$ is much smaller than the function $f(\tau)$ introduced in \cite{omidvar2017self2}.}
\label{fig:epsilon}
\end{figure}

The function $g$ is plotted in Figure~\ref{fig:epsilon}.
The following lemma, which is lemma~20 in \cite{omidvar2017self2},  gives a lower bound and an upper bound for the probability that an arbitrary neighborhood with the size of a radical region is a radical region in the initial configuration.

\begin{lemma*}[\cite{omidvar2017self2}] \label{lemma:radical_region_prob} 
Let $\epsilon'$ and $\epsilon$ be positive constants. There exist positive constants $c_l$ and $c_u$ which depend only on $\tau$ such that in the initial configuration, an arbitrary neighborhood with radius $(1+\epsilon')w$ is a radical region with probability $p_{\epsilon'}$,
\begin{align*}
c_{l}{2^{-[1-H(\tau'')](1+\epsilon')^2{\size}-o({\size})} \le p_{\epsilon'} \le c_{u} 2^{-[1-H(\tau'')](1+\epsilon')^2{\size}+o({\size})}},
\end{align*}
where $\tau'' = (\lfloor \hat{\tau}(1+\epsilon')^2{\size} \rfloor - 1) / (1+\epsilon')^2{\size} $, $\hat{\tau} = (1-{1}/{(\tau {\size}^{1/2-\epsilon})})\tau$, and $H$ is the binary entropy function.
\end{lemma*}

\subsection{The Concentration Bound} \label{Sec:Concentration}
The concentration bound we present in this section is one of the main results of~\cite{omidvar2018shape}.

\subsection{Notational conventions}
Let $t_1,t_2,...$ be independent random variables, each with distribution $F$. For a vertex or vector $v = (v(1),...,v(d))$ we shall use the $l_\infty$, $l_1$ and the $l_2$ norm. These are denoted by
\begin{align*}
||v||_\infty = \max_{1\le i\le d} |v(i)|, \ \ ||v||_1 = \sum_{i=1}^d |v(i)| \mbox{ and } ||v|| = \left\{ \sum_{i=1}^d (v(i))^2 \right\}^{1/2},
\end{align*}
respectively. $\lfloor a \rfloor$ ($\lceil a \rceil$) is the largest (smallest) integer $\le a \  (\ge a), a \wedge b = \min(a,b), a \vee b = \max(a,b)$. 

\subsection{Setup and definitions}
 \

\textit{Flipping time.} Let us call the waiting time of an unstable particle $i$ the \textit{flipping time} and denote it by $t_i$. 
Although we work with exponentially distributed waiting times, the results of this section holds for any distribution that satisfies (\ref{eq:1.1}), (\ref{eq:1.2}), and (\ref{eq:1.14}).

\

\textit{Affected* node and block}: We define an \textit{affected* node} as a node such that regardless of the states of its neighbors a $\theta$-particle on it is labeled as unstable. An \textit{affected* block} is a $w$-block such that all the nodes on it are affected*. 

\

\textit{Path and first passage time.} 
 Let a path $r$ be a set of particles such that they can flip their states in a sequence $v_1,v_2,...,v_k$. Let
\begin{align*}
T(r) = \sum_{i=1}^{k} t_i.
\end{align*}

 Also let $\mathcal{N}$ be a neighborhood containing at least one affected* block, and $u$ an arbitrary node on $G_w$. Then, we define the \textit{first passage time} from $\mathcal{N}$ to $u$ as
\begin{align*}
T(\mathcal{N},u) = \inf_{r\in \mathcal{P}} \{T(r) \},
\end{align*}
where $\mathcal{P}$ is the set of possible sequences of flips started by the affected* block in $\mathcal{N}$ that will lead to an affected node at $u$, when we assume that in the initial configuration the entire $G_w$ is a region of expansion and there are no affected nodes on $G_w$ (with these assumptions $\mathcal{P}$ is always nonempty and each particle will only flip once).

Finally, we define the first passage time or distance from $0$ to $u$ as
\begin{align*} 
a_{0,u} := T(\mathbf{0^*},u), 
\end{align*}
where $\mathbf{0^*}$ denotes an affected* block centered at the origin.


\begin{theorem*}[\cite{omidvar2018shape}]\label{Thrm:Theorem_Concentration}
Let $c>0$ be any constant and let $u$ be any node on $G_w$ whose $l_2$-distance from the origin is at least $2^{c\size}$.
There exist a constant $c'>0$ (independent of $\size$) such that for $\size$ sufficiently large, when $G_w$ is a region of expansion, and there are no affected nodes on $G_w$,  for all $\lambda\le ||u||$,  we have
\begin{align} \label{eq:1.15}
P\left\{\left|a_{0,u}-\mathbb{E}[a_{0,u}] \right| \ge \lambda    \right\} \le e^{-c'\lambda/\sqrt{||u||\log ||u||}}.
\end{align}

\end{theorem*}



The proof of the above theorem can be found in \cite{omidvar2018evolution}.

\subsection{The Shape Theorem} \label{subsec:the_shape}

We work with flipping times with i.i.d. distributions $F$ that satisfy (\ref{eq:1.1}), (\ref{eq:1.2}), and (\ref{eq:1.14}). We consider our graph $G_w$ on an infinite integer lattice. 
We work with the obvious probability space defined by our process on the lattice (i.e., product space for the initial configuration and the waiting times). We introduce the following definition from Tessera~\cite{tessera2014speed} with some modifications. 


\begin{definition}[Strong Asymptotic Geodesicity  (SAG) for $\mathbb{E}{[}a{]}$]: Let $Q: \mathbb{R}_+ \rightarrow \mathbb{R}_+$ be an increasing function such that 
\begin{align*}
\lim_{\alpha \rightarrow \infty}Q(\alpha) = \infty.
\end{align*}
Let $\mathcal{N}$ be a neighborhood 
and let $c'_1\in(0,c_1)$ be a constant. $\mathbb{E}[a]$ is called $SAG(Q)$ in $\mathcal{N}$ when for all integer $m\ge 1$, for all $x,y \in \mathcal{N}$ such that $ \mathbb{E}[a_{x,y}]/m \ge 2^{c'_1\size}$, there exists a sequence $x=x_0,....,x_m=y$ in $\mathcal{N}$ satisfying, for all $0\le i \le m-1$,
\begin{align}\label{eq:Tess_EQSAG1}  
\alpha\left(1 - \frac{1}{Q(\alpha)} \right) \le E[a_{x_i,x_{i+1}}] \le \alpha\left(1 + \frac{1}{Q(\alpha)} \right),
\end{align}
where $\alpha	= \mathbb{E}[a_{x,y}]/m$; and for $ r \ge 2^{c'_1\size}$, 
\begin{align}\label{eq:Tess_EQSAG2} 
\bar{A}\left(0,\left(1+\frac{1}{Q(r)} \right)r \right) \subset [\bar{A}(0,r)]_{\frac{6r}{Q(r)}},
\end{align}
where $\bar{A}(0,r) = \left\{x \  | \  \mathbb{E}[a_{0,x}]  \le r \right\}$ and $[\bar{A}]_t$ denotes the $t$-neighborhood of the subset $\bar{A}$ with respect to $\mathbb{E}[a]$. 
\end{definition}

\begin{proposition*}[\cite{omidvar2018shape}] \label{Prop:Tess_avg_SAG}
There exists a constant $c>0$ such that for sufficiently large $\size$, $\mathbb{E}[a]$ is SAG($Q$) in $\mathcal{N}_\rho$ where $\rho=2^{c_2\size}$ and
\begin{align} 
Q(\alpha) = \frac{\alpha^{1/2}}{\size^c\log^{3/2} \alpha}.
\end{align} 
\end{proposition*} \label{Corr:New_SAG}
The following corollary follows from the proof of the above proposition and is given without proof.
\begin{corollary*}
Let $\mathcal{N}$ be a neighborhood 
and let $c'_1\in(0,c_1)$ be a constant. For all $x,y \in \mathcal{N}$ such that $ a_{x,y}/m \ge 2^{c'_1\size}$, there exists a sequence $x=x_0,....,x_m=y$ in $\mathcal{N}$ satisfying, for all $0\le i \le m-1$,
\begin{align}\label{eq:Tess_EQSAG1}  
\alpha\left(1 - \frac{1}{Q(\alpha)} \right) \le a_{x_i,x_{i+1}} \le \alpha\left(1 + \frac{1}{Q(\alpha)} \right),
\end{align}
where $\alpha	=a_{x,y}/m$; and for $ r \ge 2^{c'_1\size}$.
\end{corollary*}

\begin{proposition*}[\cite{omidvar2018shape}]\label{Prop:Tess1}
Let $2^{c_1\size} \le n \le  2^{c_2\size}$ and $$A'_F(0,n) = \{ x \given a_{0,x} \le n  \}.$$
For sufficiently large $\size$,  there exists a norm $l_*$ on $\mathbb{R}^2$, and $c>0$, such that almost surely,
\begin{align}\label{eq:Tess1}
B_{l_*}(0,n-\size^c n^{1/2}\log^{3/2} n)\cap \mathbb{Z}^2 \subset A'_{F}(0,n)  \\ \subset B_{l_*}(0,n+\size^c n^{1/2}\log^{3/2} n). \nonumber
\end{align}
\end{proposition*}


Let $\bar{d}_1$ and $\bar{d}_2$ be the metrics defined by the expected value of first passage times in the first and second intermediate FPPs when we choose the nodes to be on $G_w$ defined on a two dimensional integer lattice. Let $x,y$ denote two arbitrary nodes on $G_w$. Assuming that $G_w$ is a region of expansion and there are no affected nodes on $G_w$, it is easy to see that there exists $c,c' >0$ such that for sufficiently large $\size$ we can write
\begin{align}\label{eq:bilips}
 d_{l_2}(x,y)/\size^c \le \bar{d}_1(0,\xi_1||x-y||_1/\size^{c'})  \le \mathbb{E}[a_{x,y}] \le \bar{d}_2(x,y) \le \size^c d_{l_2}(x,y).
\end{align}

\section{Proof of the Main Theorem}\label{Sec:Proof}
Without loss of generality, we assume that the set of nodes of $G_w$ is a subset of the nodes on $\mathbb{Z}^2$ and we work with the obvious probability space. 
We first define the \textit{expandable region}; to do so we need the following lemma from \cite{omidvar2018shape}.

\begin{figure}[!t] 
\centering
\includegraphics[width=2.4in]{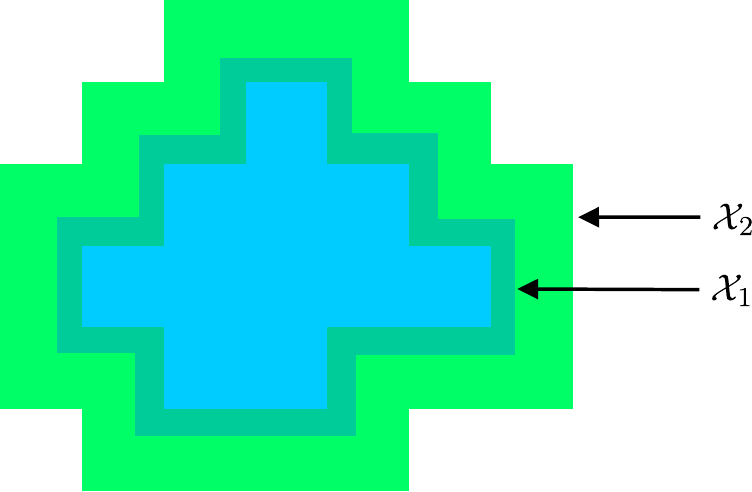}
\caption{Expandable region.  $\mathcal{X}_1$ is called a $\theta$-\textit{expandable region}  whenever there exists a set of flips of $\theta$-particles inside $\mathcal{X}_1$ leading to a $\theta$-affected node in $\mathcal{X}_2$.}
\label{fig:expandable}
\end{figure}

\begin{lemma*}[\cite{omidvar2018shape}]\label{lemma:bad_cluster}
Let $c$ be an arbitrary positive constant. W.h.p. there are no clusters of bad blocks with radius greater than $\csize$ in a neighborhood with radius $\rho = O(2^{cN})$ in the initial configuration.
\end{lemma*}

Now re-normalize $G_w$ into $\msize$-blocks and consider the union of particles inside a cluster of bad $\msize$-blocks and the set of particles outside the cluster whose $l_\infty$ distance to at least one node in the cluster is less than or equal to $\msize/4$. We denote this set by $\mathcal{X}_1$.  Note that for sufficiently large $\size$, the probability of having a bad $\msize$-block is below the critical probability of percolation, and each $\msize$-block is a bad $\msize$-block independently of the others, hence by lemma~\ref{lemma:bad_cluster}, w.h.p. there is no cluster of bad $\size$-blocks with radius larger than $\csize$ in a neighborhood with exponential size in $\size$ on $G_w$. Also, consider the set of all the particles outside $\mathcal{X}_1$ whose $l_\infty$ distance to at least one particle in $\mathcal{X}_1$ is less than or equal to $\msize/4$ and denote it by $\mathcal{X}_2$.

\  

 \textit{Expandable region.}  $\mathcal{X}_1$ is called a $\theta$-\textit{expandable region}  whenever there exists a set of flips of $\theta$-particles inside $\mathcal{X}_1$ leading to a $\theta$-affected node in $\mathcal{X}_2$ (see Figure~\ref{fig:expandable}). It is noted that if $\mathcal{X}_1$ is not an expandable region, the possible spread of $\theta$-affected nodes started in it will die out before reaching $\mathcal{X}_2$. The center of an expandable region is the node at the center of the smallest neighborhood that contains the expandable region.

\

The following lemma is lemma~13 in \cite{omidvar2018shape}. 
\begin{lemma*}[\cite{omidvar2018shape}]\label{lemma:pre_expandable}
Consider a good $10w$-block denoted by $\mathcal{N}$ in the initial configuration. Now assume that only $\theta$-particles can flip their states and at time $n$ there is a $\theta$-affected node $u$ at the center of $\mathcal{N}$. Then, at time $n$, w.h.p. there exists a sequence of flips that if they happen while the particles outside $\mathcal{N}$ maintain their configuration, there will be a $\theta$-affected $w$-block inside $\mathcal{N}$.
\end{lemma*}

We now want to argue about the distance of the closest expandable region to the origin. The following lemma shows how far the closest expandable region to the origin can be. We show this by establishing a relationship between radical regions and expandable regions. The following lemma which is lemma~5 in \cite{omidvar2018shape} exploits the fact that the closest radical region to the origin is also an expandable region w.h.p. (note that an expandable radical region is defined differently from an expandable region.)

\begin{lemma*}[\cite{omidvar2018shape}]\label{lemma:rho}
Let $\epsilon > 0$. W.h.p. the $l_*$-distance of the origin from the node  at the center of the closest expandable region in the initial configuration is at most
\begin{align*}
\rho =2^{0.5(1-H(\tau)+\epsilon)(1+\epsilon')^2N}.
\end{align*}
\end{lemma*}

The following lemma, which is lemma~6 in \cite{omidvar2018shape}, shows that in the initial configuration w.h.p. there is no part of an expandable region in an annulus around the origin whose width is $\rho'$.

\begin{lemma*}[\cite{omidvar2018shape}]\label{lemma:rhop}
Let $\epsilon > 0$. W.h.p. there is no node that belongs to an expandable region in \begin{align*}
B_{l_*}{(0,\rho+\rho')}\setminus B_{l_*}{(0,\rho)},
\end{align*}
 for
\begin{align*}
\rho =2^{0.5(1-H(\tau)+\epsilon/2)(1+\epsilon')^2\size}, \\
\rho' = 2^{(1-H(\tau)-\epsilon)(1-0.5(1+\epsilon')^2)\size + o(\size)}.
\end{align*} 
\end{lemma*}


The following lemma, which is lemma~14 in \cite{omidvar2018shape}, shows that w.h.p. an expandable region can lead to the formation of a $\theta$-affected $w$-block.

\begin{lemma*}[\cite{omidvar2018shape}]\label{lemma:expandable_to_m'}
W.h.p. there exists a sequence of possible flips in $\mathcal{X}_1 \cup \mathcal{X}_2$ that can lead to a $\theta$-affected $w$-block centered in $\mathcal{X}_1$. 
\end{lemma*}

We are now ready to give the proof of our main result. 

\begin{proof}[Theorem~\ref{Thrm:main_thrm}]
 We first show that the size of the monochromatic region is at least exponential in $\size$ w.h.p. Let $\epsilon > 0$, and
\begin{align*}
a(\tau) = \left(1-H(\tau)-\epsilon\right) \left(2-(1+\epsilon')^2\right),
\end{align*}
where $\epsilon' > g(\tau)$.
Let $n^* = 2^{(a(\tau) + \epsilon)\size}$. We wish to show that for all $n \ge n^*$,
\begin{align*}
M_n \ge 2^{a(\tau)\size} \mbox{ w.h.p.}
\end{align*}
By lemma~\ref{lemma:bad_cluster}, w.h.p. there is no cluster of bad blocks with radius larger than $\csize$ in a neighborhood with radius $2^{\size}$ centered at the origin in the initial configuration (event $A_0$). 

Let
\begin{align*}
\rho =2^{0.5(1-H(\tau)+\epsilon/2)(1+\epsilon')^2\size}, \\
\rho' = 2^{(1-H(\tau)-\epsilon)(1-0.5(1+\epsilon')^2)\size + 2\log_2 \size},\\
\rho'' = 2^{(1-H(\tau)+\epsilon)\left((1+\epsilon')^2-1\right)\size},
\end{align*} 


We let $\size$ be sufficiently large so that there exists a norm $l_*$ and $C>0$ such that (\ref{eq:Tess1}) in Proposition~\ref{Prop:Tess1} is satisfied for $n = \rho'^{1/3}$, $\rho'/\size > w^3$. Let  
\begin{align*}
L= \left\lceil \frac{\rho}{\rho'/4}\right\rceil.
\end{align*}

Let the closest expandable region to the origin in the $l_*$ norm be a $\theta$-expandable region and let $\mathcal{N}_{\rho'/4}$ be a neighborhood at the origin with radius $\rho'/4$, now let
\begin{align*}
T(\rho'/4) = \inf\left\{n \  \given[\Big] \  \exists \mbox{ a $\bar{\theta}$-affected node in } \mathcal{N}_{\rho'/4} \right\}, \\
A = \left\{ \mbox{The origin is contained in a firewall of radius $\rho'/\size$ before } T(\rho'/4)  \right\}.
\end{align*} 

We now want to show
\begin{align} \label{eq:P_A}
A \mbox{ occurs w.h.p.}
\end{align}
To show this, we condition this event on a few events that occur w.h.p. and argue that since the conditional probability occurs w.h.p., this event also occurs w.h.p. 


 Let $X$ denote the $l_*$-distance from the origin to the closest node in an expandable region and let
\begin{align*}
&A_1 = \left\{ \mbox{$X\le \rho$, at $n=0$}  \right\},  \\
&A_2 = \left\{ \nexists \mbox{ a $\bar{\theta}$-expandable region in } B_{l_*}(0,X+\rho')\setminus B_{l_*}(0,X) \mbox{ at $n=0$} \right\}.
\end{align*}

\begin{figure}[!t] 
\centering
\includegraphics[width=4in]{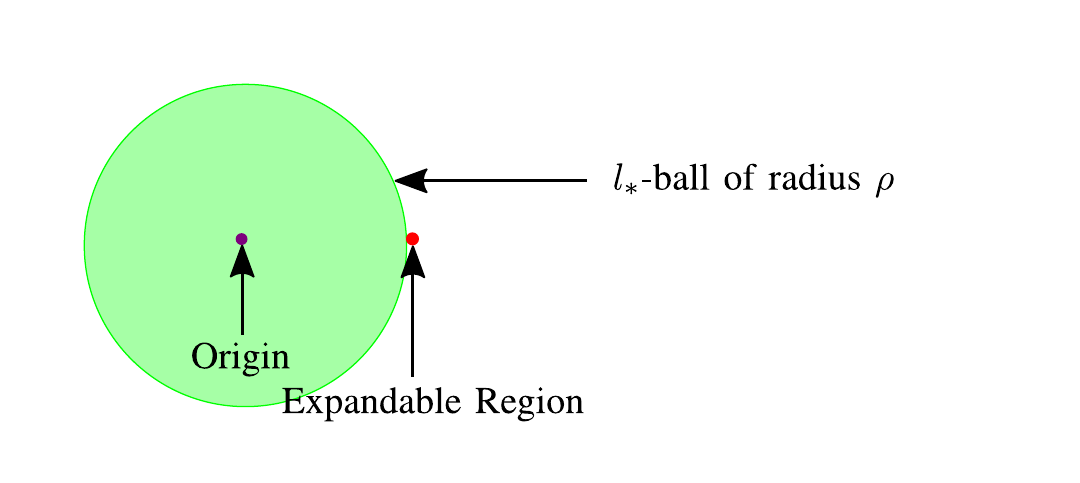}
\caption{W.h.p. the closest expandable region to the origin will not be farther than $\rho$ with respect to the $l_*$-norm. } 
\label{fig:closest_expandable} 
\end{figure}

Consider an $l_*$-ball of radius $\rho$. According to lemma~\ref{lemma:rho}, w.h.p. there is an expandable region in this ball (see Figure~\ref{fig:closest_expandable}). This implies
\begin{align*}
A_1 \mbox{ occurs w.h.p.}
\end{align*}
Using the fact that the existence of a $\theta$-expandable region  in $B_{l_*}(0,X+\rho')\setminus B_{l_*}(0,\rho)$ can only increase the probability of event $A_2$ (since they are both increasing events in the change of a $\theta$-particle to a $\bar{\theta}$-particle, by an application of FKG inequality \cite{fortuin1971correlation} for the initial configuration they are positively correlated), and the fact that conditional on event $A_1$, event $A_2$ would have the smallest probability when $X=\rho$, we let
\begin{align*}
&A'_2 = \left\{ \nexists \mbox{ a $\bar{\theta}$-expandable region in } B_{l_*}(0,\rho+\rho')\setminus B_{l_*}(0,\rho) \mbox{ at $n=0$} \right\},
\end{align*}
and by an application of FKG inequality \cite{fortuin1971correlation} for the initial configuration we have
\begin{align*}
P(A_2) \ge P\left(A_2 \given[\Big] A_1\right) P(A_1) \ge P\left(A_2 \given[\Big] X = \rho\right) P(A_1) \ge P(A'_2)P(A_1).
\end{align*}
Using lemma~\ref{lemma:rhop}, event $A'_2$ occurs w.h.p., hence we have
\begin{align*}
A_2 \mbox{ occurs w.h.p.}
\end{align*}

Now consider the line segment from the center of the closest expandable region to the origin. Let  $\mathcal{N}$ denote the set of particles such that their $l_*$-distances from at least one point on the line segment is less than or equal to $2\size^{c'}\rho'$ where $c'$ is a positive constant. Let
\begin{align*}
\begin{split}
A_3 = \left\{ \nexists \mbox{ a $\bar{\theta}$-affected node in } \mathcal{N} \mbox{ and it is a region} \right. \\
\left. \mbox{ of expansion at $n=0$ }  \right\}.
\end{split}
\end{align*}

Since event $A_3$ has the smallest probability when $X=\rho$, with an application of lemma~\ref{lemma:unstableprob} and lemma~\ref{lemma:monoch_spread_1}, we can conclude that 
 \begin{align*}
A_3 \mbox{ occurs w.h.p.}
\end{align*}



 \begin{figure}[!t] 
\centering
\includegraphics[width=4in]{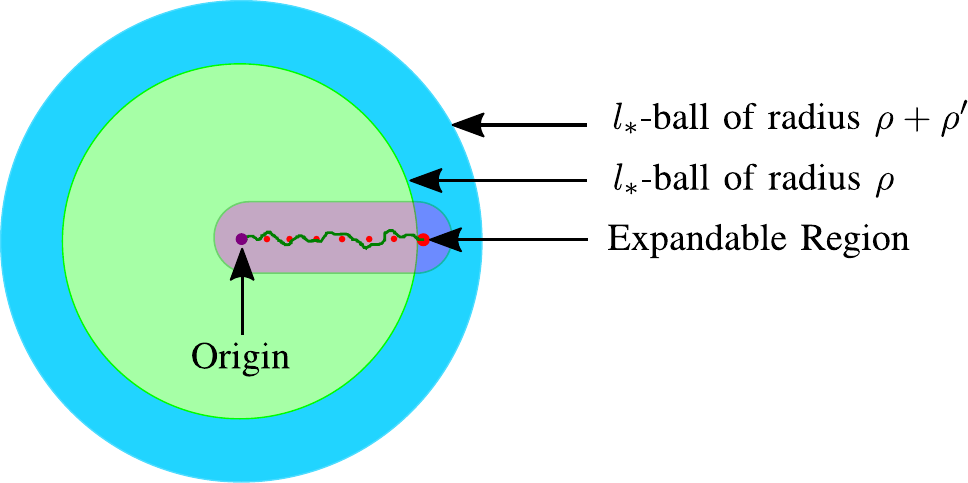}
\caption{The spread of affected nodes from the expandable region towards the origin. } 
\label{fig:gradual_growth2}
\end{figure} 
We are now going to consider the spread of the affected nodes from the expandable region towards the origin in $\rho'/4$ time intervals (see Figure~\ref{fig:gradual_growth}). 

To consider the growth of the $\theta$-affected nodes 
we first notice that using lemma~\ref{lemma:expandable_to_m'}, w.h.p. there exists a sequence of less than $\size^4$ flips that can create a $\theta$-affected $w$-block centered at the center of the expandable region. Let us denote this event by $A_4$. Hence we have
\begin{align*}
A_4 \mbox{ occurs w.h.p. }
\end{align*} 
Let $T_{\size}$ denote the time it takes until $\size^4$ flips occur one by one. Let $A'_5 = \{T_{\size} < \rho'^{1/3} \}$.  Standard concentration bounds imply that there exist $c>0$ such that this event occurs with probability at least $1-\exp(-c\rho'^{1/3})$. Let $A_5$ denote the event that the time that it takes until we have a $\theta$-affected block at the center of the expandable region is less than $\rho'^{1/3}$. We have
\begin{align*}
P(A_5)  \ge P\left(A'_5\right) \ge  1-\exp\left(-c\rho'^{1/3}\right).
\end{align*} 
Hence we have 
\begin{align*}
A_5 \mbox{ occurs w.h.p.}
\end{align*}

Let $A_6$ denote the event of having less than $\rho''$ affected nodes in the $l_*$-ball of radius $\rho$. Standard concentration bounds imply that this event also occurs w.h.p., hence we have 
\begin{align*}
A_6 \mbox{ occurs w.h.p.}
\end{align*}

Now we want to consider the spread of possible $\bar{\theta}$-affected nodes from outside of $B_{l_*}(0,\rho+\rho')$ towards the origin. Since there are at most $\rho''$ affected nodes in this ball w.h.p., if we remove all the nodes contained in the annuli in this ball that contain an affected node along with their clusters of bad blocks and also a margin of good blocks around these clusters we would have a ball of radius at least $\rho+\rho'-(\size^4+\size)\rho''$. Furthermore, we argue that since the event of having larger growths of $\bar{\theta}$-affected nodes in a given time interval and the event of having this ball being a region of expansion of type $\bar{\theta}$ are both increasing in the change of a $\bar{\theta}$-particle to a $\theta$-particle, by assuming that this ball is a region of expansion of type  $\bar{\theta}$ we would only get an upper bound for the speed of the spread of $\bar{\theta}$-affected nodes. Hence, from this point forward, to consider the spread of $\bar{\theta}$-affected nodes towards the origin we consider this ball and assume it is a region of expansion and does not contain any affected nodes. 

 Let $A_7$ denote the event that the growths of the $\bar{\theta}$-affected nodes in a time interval of $\rho'^{1/3}$ -- that conditional on $A_i$, $i=1,2,...,6$ is at most needed for the formation of the $\theta$-affected $w$-blocks in the first gradual growth -- is less than  $\rho'^{1/3}+ \size^C\rho'^{1/6}\log^{3/2}\rho' $ in the $l_*$ norm in all directions in the annulus around the origin. To show that this event occurs w.h.p. we consider the $l_*$-ball of radius $\rho - (\size^4+\size)\rho''$ described above. It follows from Proposition~\ref{Prop:Tess1} that the growths of all the $\bar{\theta}$-affected nodes in this ball will be less than $\rho'^{1/3}+ \size^C\rho'^{1/6}\log^{3/2}\rho'$ hence we can conclude that
\begin{align*}
A_7  \mbox{ occurs w.h.p. }
\end{align*}

Now let $A_8$ denote the event that the origin is contained in a $w$-block of $\theta$-affected nodes before there are any $\bar{\theta}$-affected nodes in an $l_*$-ball with radius $\rho'/2$ around the origin. To show that this event occurs w.h.p. we consider $L$ equally distanced points $\{p_1,p_2,...,p_L\}$ on the line segment connecting the center of the expandable region to the origin.

We now consider
time intervals of size $\rho'/4$ and argue that using Corollary~\ref{Corr:New_SAG} in every one of these time intervals $i\in[L]$ there will be a $\theta$-affected node with an $l_*$-distance of at most $O(\size^c\sqrt{\rho})$ from $p_i$ where $c>0$ is a constant. Also, using lemma~\ref{lemma:pre_expandable} there will also be a $\theta$-affected block with an $l_*$-distance of at most $O(\size^c\sqrt{\rho})$ from $p_i$.


We argue that not having $\theta$-affected nodes in $\mathcal{N}'$ and having smaller growths of $\bar{\theta}$ in a given time interval in this neighborhood are both increasing events in the change of a $\bar{\theta}$-particle to a $\theta$-particle hence using the FKG inequality these events are positively correlated. 


Now we also need to show that the possible growths of $\bar{\theta}$-affected nodes started from outside of $B_{l_*}(0,\rho+\rho'-w)$ are not going to interfere with any of these growths and will not reach the $B_{l_*}(0,\rho'/2)$ before having the origin $\theta$-affected (see Figure~\ref{fig:gradual_growth2}). To see this, since we are conditioning on events $A_i$, $i=1,2,...,7$, 
 it suffices to show that the growths of $\bar{\theta}$-affected nodes started from outside a $B_{l_*}(0,\rho+\rho'-\rho'^{1/3}-\size^c\rho''-\size)$ which does not contain any affected nodes and is assumed to be a region of expansion in every time interval of size $\rho'/4$ is at most $\rho'/4+\size^C(\rho)^{1/2}\log^{3/2}$ in the $l_*$-norm, which guarantees that not only these growths will not interfere with the growths of $\theta$-affected nodes but also they will not reach  $B_{l_*}(0,\rho'/2)$ before having the origin contained in a $\theta$-affected $w$-block. Since Corollary~\ref{Corr:New_SAG} applies to every pair of nodes in this region it follows from a similar argument for the spread of $\theta$-affected nodes, that this event also occurs w.h.p. Hence we can conclude that 
 \begin{align*}
 A_8 \mbox{ occurs w.h.p.}
 \end{align*}

This also implies that by the time the origin has become a $\theta$-affected node, the $\bar{\theta}$-affected nodes are still in $l_*$-distance of more than $\rho'/2$ from the origin w.h.p. We also note that if the above event occurs, using lemma~\ref{lemma:pre_expandable} there is also a $\theta$-affected $w$-block in an $l_*$ distance of $o(\size)$ to the origin.


Now let $r$ be proportional to $\rho'/\size$. Let us denote the event that the time it takes until a number of affected nodes equal to the number of all the particles in a firewall with radius $r$ centered at the origin and a line of width $2\sqrt{\size}$ from the origin to the firewall and also a line of width $2\sqrt{\size}$ from the closest $\theta$-affected $w$-block to the origin, make a flip one by one being smaller than $\rho'/4$ by $A'_9$  (see Figure~\ref{fig:firewall_formation}).  Standard concentration bounds imply $A'_9 \mbox{ occurs w.h.p.}$

Let $A_9$ denote the event that this firewall is formed in a time interval smaller than $\rho'/4$. We have $P(A_9) \le P(A'_9)$ and since $A'_9$ occurs w.h.p. we have
\begin{align*}
A_9 \mbox{ occurs w.h.p.}
\end{align*} 

With a similar argument for event $A_8$, w.h.p. the growth of all the possible $\bar{\theta}$ affected nodes will be less than $\rho'/3$ for this interval (event $A_{10}$). Hence, we have
\begin{align*}
A_{10} \mbox{ occurs w.h.p.}
\end{align*}

 Finally we can write
 \begin{align*}
 P(A) \ge P\left(A\given[\Big] A_0,A_1,...,A_{10}\right)P(A_0\cap A_1 \cap ... \cap A_{10}),
 \end{align*}
hence, we have that 
\begin{align*}
A \mbox{ occurs w.h.p.}
\end{align*}
Now, using lemma~\ref{lemma:monoch_spread_1} w.h.p. the interior of the firewall is a region of expansion in the initial configuration and since w.h.p. only $\theta$-affected nodes have reached this region by the time of the formation of the firewall, it is still a region of expansion for the state $\bar{\theta}$. Now, since the sum of the time for the gradual growths, formation of the firewall, and the time that it takes until the interior of the firewall becomes monochromatic (by a standard concentration bound) is less than $n^*$ w.h.p., for all $n\ge n^*$ it will be monochromatic w.h.p. and this proves the lower bound.

\begin{figure}[!t] 
\centering
\includegraphics[width=1.6in]{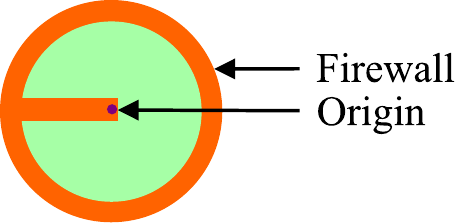}
\caption{Formation of a firewall around the origin.} 
\label{fig:firewall_formation}
\end{figure}


Next, we show the corresponding upper bound. Let
\begin{align*} 
b(\tau)= \left(1+\epsilon'\right)^2(1-H(\tau)+\epsilon).
\end{align*}
Consider four neighborhoods with radius $\size(\rho+\rho')$ such that each of them share the origin as a different corner node. Divide the union of these neighborhoods into neighborhoods of radius $\rho+\rho'$ in an arbitrary way and consider the nodes at the center of each of these neighborhoods. Now using the above result we have that for $n\ge n^*$, w.h.p., all these central nodes will have a monochromatic region of size at least $2^{a(\tau)\size}$. Also it is easy to see that for $n\ge n^*$, w.h.p. all the four neighborhoods defined above will have particles with exponentially large monochromatic regions of both states. This implies that for all $n\ge n^*$ the size of the monochromatic region of the origin is at most $4\size^2(\rho+\rho')^2$.
Finally let
\begin{align*}
A_w = \left\{2^{a(\tau){\size}} \le M_n \le 2^{b(\tau){\size}} \mbox{ on } G_w \mbox{ for } n\ge n^* \right\}.
\end{align*}
We have that $P(A^C_w) = o(w^{-2})$. The result now follows from the fact that $\sum_{w=1}^\infty P(A^C_w) < \infty$.
\end{proof}

\subsection{Extension to $\tau>1/2$}\label{Subsec:Extension}
We call  \textit{super-unstable particles} the unstable particles that can potentially become stable once they flip their state.
While for $\tau<1/2$ unstable particles  can always  become stable by flipping their state,  for $\tau>1/2$ this is only true for the super-unstable particles. 
It follows that for  $\tau > 1/2$ super-unstable particles act  in the same way as unstable particles do for $\tau < 1/2$. 


We let $\bar{\tau} = 1 - \tau + 2/{\size}$. A \textit{super-unstable particle} of type $\theta$ is a particle for which $W < \bar{\tau}{\size}$ where $W$ is the number of $\theta$ particles in its neighborhood. The reason for adding the term $2/{\size}$ in the definition is to account for the strict inequality that is needed for being unstable and the flip of the particle at the center of the neighborhood which adds one particle of its type to the neighborhood. 
A \textit{super-radical region} is a neighborhood $\mathcal{N}_S$ of radius $S=(1+\epsilon')w$ such that $W_S <  \bar{\tau}'(1+\epsilon')^2{\size} $, where    $\epsilon \in (0,1/2)$ and
\begin{equation*}
\bar{\tau}' = \left(1-\frac{1}{\bar{\tau} {\size}^{1/2-\epsilon}}\right)\bar{\tau}. 
\end{equation*}

By replacing $\tau$ with $\bar{\tau}$, ``unstable particle'' with ``super-unstable particle''  and   ``radical region'' with  ``super-radical region,''   it can be checked that all proofs extend to the interval $1/2<\tau<1-\tau^*$.

\section*{Acknowledgment}
The authors  thank Prof. Jason Schweinsberg  of the Mathematics Department of  University of California at San Diego for providing invaluable feedback and suggestions on earlier drafts of this paper.

\printbibliography


\end{document}